\documentclass[12pt]{article}
\usepackage[dvips]{graphicx}
\usepackage[cp1251]{inputenc}
\usepackage[english]{babel}
\usepackage{amssymb}
\usepackage{amsmath}
\usepackage{amsthm}
\usepackage{amsbsy}
\usepackage{amsfonts}
\fontsize{16}{20}
\numberwithin{equation}{section}

\begin{document}
\begin{center}
\begin{Large}
{\bf Asymptotic and numerical studies of resonant tunneling in 2D
quantum waveguides of variable cross-section}
\end{Large}

\vspace{4ex}
\begin{large}
Lev Baskin, Muaed Kabardov, Pekka Neittaanm\"aki, Boris
Plamenevskii, and Oleg Sarafanov
\end{large}
\footnote{Supported by grant RFBR-09-01-00191-a }
\end{center}
\begin{abstract}
A waveguide coincides with a strip having two narrows of diameter
$\varepsilon$. Electron motion is described by the Helmholtz
equation with Dirichlet boundary condition. The part of waveguide
between the narrows plays the role of resonator and there can occur
electron resonant tunneling. This phenomenon consists in the fact
that, for an electron with energy $E$, the probability $T(E)$ to
pass from one part of the waveguide to the other part through the
resonator has a sharp peak at $E=E_{res}$, where $E_{res}$ denotes a
"resonant" energy. In the present paper, we compare the asymptotics
of $E_{res}=E_{res}(\varepsilon)$ and $T(E)=T(E, \varepsilon)$ as
$\varepsilon \to 0$ with the corresponding numerical results
obtained by approximate computing the waveguide scattering matrix.
We show that there exists a band of $\varepsilon$ where the
asymptotics and numerical results are in close agreement. The
numerical calculations become inefficient as $\varepsilon$
decreases; however, at such a condition the asymptotics remains
reliable. On the other hand, the asymptotics gives way to the
numerical method as $\varepsilon$ increases; in fact, for wide
narrows the resonant tunneling vanishes by itself.

Though, in the present paper,  we consider only a 2D waveguide, the
applicability of the methods goes far beyond the above simplest
model. In particular, the same approach will work for asymptotic and
numerical analysis of resonant tunneling in 3D quantum waveguides.
\end{abstract}
\section{Introduction}\label{s1}
As an electron propagates in a quantum waveguide of variable
cross-section, the waveguide narrows play the role of effective
potential barriers for the longitudinal motion. The part of the
waveguide between two narrows becomes a "resonator"\,, and there can
arise resonant tunneling. It consists of the fact that, for an
electron with energy $E$, the probability $T(E)$ to pass from one
part of the waveguide to the other through the resonator has a sharp
peak at $E=E_{res}$, where $E_{res}$ denotes a resonant energy.
There are prospects for building a new class of nanosize electronics
elements (transistors, electron energy monochromators, key devices)
based on the phenomenon of resonant tunneling. To analyze their
operation, it is important to know $E_{res}$, the height of the
resonant peak, the behavior of $T(E)$ for $E$ close to $E_{res}$,
etc.

In \cite{BNPS}, electron propagation was considered in a 3D
waveguide with two cylindrical outlets to infinity and two narrows
of small diameter $\varepsilon_1$ and $\varepsilon_2$. The electron
motion was described by the Helmholtz equation with Dirichlet
boundary condition, radiation condition, and a wave number $k$
between the first and the second thresholds. For the aforementioned
characteristics of resonant tunneling, there were obtained
asymptotics as $\varepsilon_1, \varepsilon_2 \to 0$. The asymptotic
formulas provide mainly a qualitative picture. In the present paper,
we show that, being supplemented by some computations, the
asymptotics can tell a useful quantitative information as well.
Though the paper continues the studies in \cite{BNPS}, nevertheless
it is practically self-contained; let us explain its goal in detail.

The asymptotic formulas in \cite{BNPS} include several unknown
constant coefficients, which can be found by solving some boundary
value problems independent of $\varepsilon_1$ and $\varepsilon_2$.
Here, in a model situation, we
 calculate approximately such coefficients, which enables us to take
the asymptotics as numerical values of resonant tunneling
characteristics for sufficiently small $\varepsilon_1$ and
$\varepsilon_2$. This leads to the question which $\varepsilon_1$,
$\varepsilon_2$ could be considered as "sufficiently small"\,; in
other words, where does the asymptotics work in a proper way? Though
there is no universal answer for such a question, some examples give
a grasp of what should be expected in analogous cases. To this end
we calculate (also approximately) the scattering matrix and then
compare the results obtained by the asymptotic and computational
methods independently of one another. Generally, it can be predicted
that numerical calculations will become inefficient as the narrow
diameters decrease and the resonant peak turns out to be "too
sharp"\,; however, at such a condition the asymptotics should become
more reliable. On the other hand, the asymptotics will give way to
the numerical method as the narrow diameters increase; in fact, for
wide narrows the resonant tunneling  would vanish by itself. We
observe these phenomena and show that there exists a band of the
diameters, where the asymptotic and numerical approaches give
compatible results.

In the present paper, we consider a 2D waveguide that is a strip
with two narrows of the same diameter $\varepsilon$ (see Fig. \ref{Fig.G_e}).
For 2D waveguides, the asymptotics of resonant tunneling
characteristics (as $\varepsilon \to  0$) are published here for the
first time; however, we do not prove the formulas in the paper. The
reader could obtain the needed proofs by modifying arguments in
\cite{BNPS} related to a 3D situation. Nonetheless, we analyze the
structure of asymptotics in order to explain what constants in the
asymptotic formulas have to be calculated numerically and how to do
that by solving some boundary value problems independent of
$\varepsilon$.

The paper consists of five sections. The mathematical model of the
waveguide and statement of the problem are given in section 2. The
asymptotic formulas are presented in section 3. Then, in section 4,
we list the constants to be calculated in the asymptotics, describe
the boundary value problems needed for the purpose, and present the
methods for solving the problems numerically. In the same section we
also describe a method we have used for approximate computation of
the waveguide scattering matrix. Finally, section 5 is devoted to
comparing basic resonant tunneling characteristics obtained in two
different ways, asymptotic and numerical, independent of one
another.

Though in the present paper we considered only a 2D waveguide, the
applicability of the methods goes far beyond the above simplest
model. In particular, the same approach will work for comparison
asymptotic and numerical analysis of resonant tunneling in 3D
quantum waveguides.
\section{Statement of the problem}\label{s2}
\begin{figure}[!htbp]
\centering
\includegraphics[scale=.6]{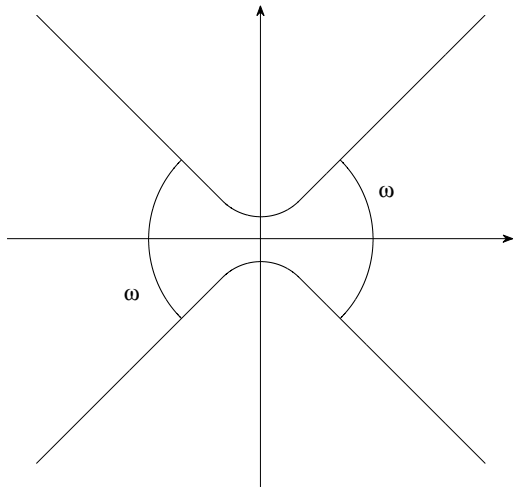}
\caption{The set $\Omega$.}
\label{Fig.Omega}
\end{figure}

To describe the domain $G(\varepsilon)$ in $\mathbb R^2$ occupied by
the waveguide, we first introduce two auxiliary domains $G$ and
$\Omega$ in $\mathbb R^2$. The domain $G$ is the strip
$$
G=\mathbb R \times D =\{(x, y)\in \mathbb R^2: x\in \mathbb
R=(-\infty, +\infty); y \in D=(-l/2, l/2)\}.
$$
Let us define $\Omega$. Denote by $K$ a double cone with vertex at
the origin $O$ that contains the $x$-axis  and is symmetric about
the coordinate axes. The set $K\cap S^1$, where $S^1$ is a unit
circle, consists of two simple arcs. Assume that $\Omega$ contains
the cone $K$ and a neighborhood of its vertex; moreover, outside a
large disk (centered at the origin) $\Omega$ coincides with $K$. The
boundary $\partial \Omega$ of $\Omega$ is supposed to be smooth (see
Fig. \ref{Fig.Omega}).

\begin{figure}[!htbp]
\vspace{-1cm}
    \centering
        \includegraphics[scale=.7]{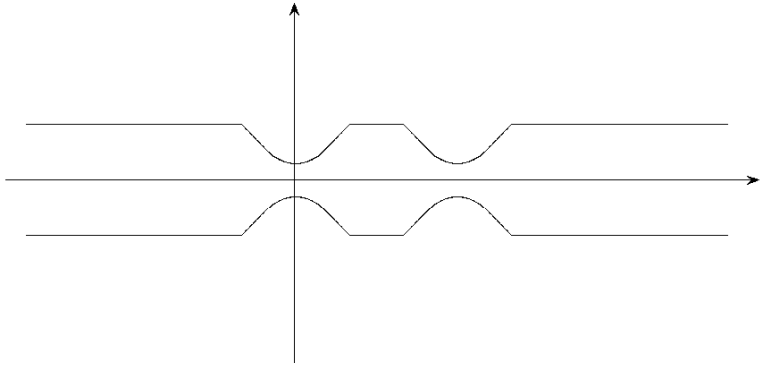}
\vspace{-1cm}
\caption{The waveguide $G(\varepsilon).$}
\label{Fig.G_e}
\end{figure}

We now turn to the waveguide $G(\varepsilon)$. Denote by $\Omega
(\varepsilon)$ the domain obtained from $\Omega$ by the contraction
with center at $O$ and coefficient $\varepsilon$. In other words,
$(x, y) \in \Omega (\varepsilon)$ if and only if $(x/\varepsilon,
y/\varepsilon) \in \Omega$. Let $K_j$ and $\Omega_j (\varepsilon)$
stand for $K$ and $\Omega (\varepsilon)$ shifted by the vector
$\mathbf{r}_j=(x_j^0, 0)$, $j=1, 2$. We assume that $|x_1^0-x_2^0|$
is sufficiently large so the distance from $\partial K_1 \cap
\partial K_2$ to $G$ is positive. We put
$$
G(\varepsilon)=G\cap \Omega_1(\varepsilon)\cap
\Omega_2(\varepsilon)
$$
(see Fig. \ref{Fig.G_e}).
The wave function of a free electron of energy $k^2$ satisfies the
boundary value problem
\begin{eqnarray}\label{2D problem}
  \Delta u(x, y)+k^2u (x, y) &=& 0, \qquad (x, y) \in G(\varepsilon),
  \\\nonumber
  u (x, y)&=& 0, \qquad (x, y) \in \partial G(\varepsilon).
\end{eqnarray}
Moreover, $u$ is subject to radiation conditions at infinity. To
formulate the conditions we need the problem
\begin{eqnarray}\label{problem on D}
  \Delta v(y)+\lambda^2v(y) &=& 0,\qquad y \in D,
  \\\nonumber
  v(-l/2)=v(l/2) &=& 0.
\end{eqnarray}
The eigenvalues $\lambda^2_q$ of this problem, where $q=1,2,\ldots$
are called the thresholds; they form the sequence $\lambda^2_q=(\pi
q/l)^2$, $q=1,2,\ldots$.  We suppose that $k^2$ in (\ref{2D
problem}) is not a threshold. Given a real $k$, there exist finitely
many linearly independent bounded wave functions. In the linear
space spanned by such functions, a basis is formed by the wave
functions subject to the radiation conditions
\begin{eqnarray}\label{um}
  u_m(x,y) &=& \begin{cases}
    e^{i\nu_mx}\Psi_m(y) + \displaystyle\sum_{j=1}^Ms_{mj}(k)\,e^{-i\nu_jx}\Psi_j(y)+O(e^{\delta x}), & x\rightarrow -\infty, \\
                \displaystyle\sum_{j=1}^Ms_{m,M+j}(k)\,e^{i\nu_jx}\Psi_j(y)+O(e^{-\delta x}), & x\rightarrow +\infty; \\
\end{cases} \\\nonumber
  u_{M+m}(x,y) &=& \begin{cases}
    \displaystyle\sum_{j=1}^Ms_{M+m,j}(k)\,e^{-i\nu_jx}\Psi_j(y)+O(e^{\delta x}), & x\rightarrow -\infty, \\
    e^{-i\nu_mx}\Psi_m(y) +\\
\hspace{1.2cm}
+\displaystyle\sum_{j=1}^Ms_{M+m,M+j}(k)\,e^{i\nu_jx}\Psi_j(y)
+O(e^{-\delta x}), & x\rightarrow +\infty. \\
\end{cases}
\end{eqnarray}
Here $M$ is the number of the thresholds satisfying $\lambda^2<k^2$;
$m=1,2,\ldots,M$; $\nu_m=\sqrt{k^2-\lambda^2_m}$; $\Psi_m$ is an
eigenfunction of the problem (\ref{problem on D}) that corresponds
to the eigenvalue $\lambda^2_m$ and is chosen so that
\begin{equation}\label{eigenfunctions}
    \Psi_m(y)=\left\{%
\begin{array}{ll}
    \sqrt{2/l\nu_m}\sin\lambda_my, & \hbox{$m$ even,} \\
    \sqrt{2/l\nu_m}\cos\lambda_my, & \hbox{$m$ odd.} \\
\end{array}%
\right.
\end{equation}
The function $ U_j(x,y)=e^{i\nu_jx}\Psi_j(y),\qquad j=1,\ldots,M$,
in the strip $G$ is a wave incoming from $-\infty$ and outgoing to
$+\infty$, while $ U_{M+j}(x,y)=e^{-i\nu_jx}\Psi_j(y),\qquad
j=1,\ldots,M,
$
is a wave going from $+\infty$ to $-\infty$. The scattering matrix
$$
S=\|s_{mj}\|_{m,j=1,\ldots,2M}
$$
is unitary. The values
\begin{equation*}
    R_m=\sum_{j=1}^M|s_{mj}|^2, \qquad \,\,\,  T_m=\sum_{j=1}^M|s_{m, \,M+j}|^2
\end{equation*}
are called the reflection and transition coefficients, relatively,
for the wave $U_m$ incoming to $G(\varepsilon)$ from $-\infty$,
$m=1, \dots, M$. (Similar definitions can be given for the wave
$U_{M+m}$ coming from $+\infty$.)

In the present work we will discuss only the case
$(\pi/l)^2<k^2<(2\pi/l)^2$, i.e., $k^2$ is between the first and the
second thresholds. Then the scattering matrix is of size $2\times2$.
We consider only the scattering of the wave incoming from $-\infty$
and denote the reflection and transition coefficients as
\begin{equation}\label{2D reflection coeff}
R=R(k, \varepsilon)=|s_{11}(k, \varepsilon)|^2, \qquad \,\,\, T=T(k,
\varepsilon)=|s_{12}(k, \varepsilon)|^2.
\end{equation}
The goal is to find a "resonant" value $k_r=k_r(\varepsilon)$ of the
parameter $k$ corresponding to the maximum of the transition
coefficient, and to describe the behavior of $T(k,\varepsilon)$ for
$k$ in a neighborhood of $k_r(\varepsilon)$ as $\varepsilon \to 0$.
\section{Outline of the asymptotics}\label{s3}
\begin{figure}[!hp]
    \centering
        \includegraphics[scale=.7]{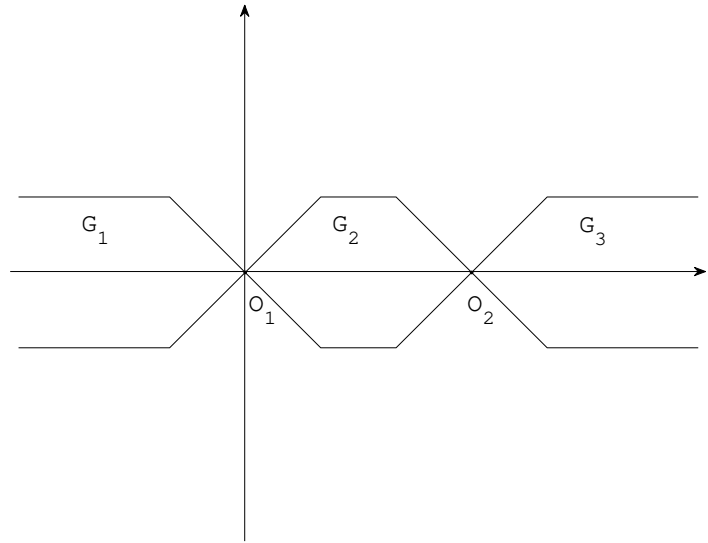}
\vspace{-1cm}
\caption{The set $G(0).$}
\label{Fig.G_0}
\end{figure}

When deriving an asymptotics of a wave function (i.e. solution of
problem (\ref{2D problem})) as $\varepsilon \rightarrow 0$, we use
the compound asymptotics method (the general theory of the method
was exposed, e.g., in \cite{MNP},  \cite{KMM}). To this end we
introduce "limit"\, boundary value problems independent of the
parameter $\varepsilon$. Put $G(0)=G\cap K_1\cap K_2$ (Fig.
\ref{Fig.G_0}); thus,  $G(0)$ consists of the three parts $G_1$,
$G_2$, and $G_3$, where $G_1$ and $G_3$ are infinite domains while
$G_2$ is a bounded resonator.

The problems
\begin{eqnarray}\label{limit problems 1}
  \Delta v(x,y)+k^2v(x,y) &=& 0,\qquad(x,y)\in G_j,
  \\\nonumber
  v(x,y) &=& 0,\qquad(x,y)\in \partial G_j,
\end{eqnarray}
where $j=1,2,3$ and $\partial G_j$ is the boundary of $G_j$, are
called the first kind limit problems. Solutions $v_1$ and $v_3$ are
subject to some radiation conditions at infinity and all three
functions $v_1$, $v_2$, $v_3$ satisfy some conditions at the corner
points. All of the conditions will be formulated as required.

Let us turn to the domains $\Omega_1$ and $\Omega_2$ (see Fig. \ref{Fig.Omega}).
Problems of the form
\begin{eqnarray}\label{limit problems 2}
  \Delta w(\xi_j,\eta_j) &=& F(\xi_j,\eta_j)\qquad\hbox{in}\;\Omega_j,
  \\\nonumber
  w(\xi_j,\eta_j) &=& 0
  \qquad\qquad\quad\hbox{on}\;\partial\Omega_j,
\end{eqnarray}
are called the second kind limit problems. We seek solutions of the
problems  satisfying
$$
w(\xi_j,\eta_j) = O\left(\rho_j^{-3\pi/\omega_j}\right)\qquad
\hbox{as}\;\rho_j\rightarrow\infty;
$$
here  $(\xi_j,\eta_j)$ are rectangular coordinates with origin at
the vertex $O_j$ of $K_j$, $\rho_j$ being the distance from
$(\xi_j,\eta_j)$ to $O_j$ and $\omega_j$ the opening of $K_j$,
$j=1,2$.

In the waveguide $G(\varepsilon)$, we consider the scattering of the
wave $U(x,y)=e^{i\nu_1x}\Psi_1(y)$ incoming from $-\infty$ (see
(\ref{eigenfunctions})). The asymptotics of the wave function is the
main technical result. Although rather cumbersome, it will lead to
much more explicit characteristics of the process. The wave function
admits the representation
\begin{eqnarray}\label{asympt of wave function}\nonumber
  &&u(x,y;\varepsilon)=\chi_{1,\, \varepsilon}(x,y)v_1(x,y;\varepsilon)+\\
  &&+\Theta(r_1)w_1(\varepsilon^{-1}x_1,\varepsilon^{-1}y_1;\varepsilon)+
     \chi_{2,\, \varepsilon}(x,y)v_2(x,y;\varepsilon)+\\\nonumber
  &&+\Theta(r_2)w_2(\varepsilon^{-1}x_2,\varepsilon^{-1}y_2;\varepsilon)+
   \chi_{3,\, \varepsilon}(x,y)v_3(x,y;\varepsilon)+R(x,y;\varepsilon).
\end{eqnarray}
Let us explain the notation and the structure of this formula. When
composing the formula, we first describe the behavior of the wave
function
 to the right of the narrows, where the wave function can be
approximated by a solution $v_3$ of the problem (\ref{limit problems
1}) in $G_3$. The solution is subject to the radiation condition
\begin{equation}\label{v3 radiation condition}
    v_3 (x,y;\varepsilon)\sim s_{12}(\varepsilon)e^{i\nu_1x}\Psi_1(y)
    \qquad\hbox{as}\;x\rightarrow+\infty,
\end{equation}
the element  $s_{12}(\varepsilon)$ of scattering matrix being yet
unknown. Problem (\ref{limit problems 1}) does not contain
$\varepsilon$, nevertheless  $v_3$ depends on the parameter because
of $s_{12}(\varepsilon)$. By $\chi_{3,\, \varepsilon}$ we denote a
cut-off function defined by
$$
\chi_{3,\, \varepsilon}(x,y)=
\left(1-\Theta(r_2/\varepsilon)\right)\mathbf{1}_{G_3}(x,y),
$$
where $r_2=\sqrt{x_2^2+y_2^2}$ and $(x_2,y_2)$ are the coordinates
of a point $(x,y)$ in the system obtained by shifting the
 origin to the point $O_2$; $\mathbf{1}_{G_3}$
is the indicator of $G_3$ (equal to 1 in $G_3$ and to 0 outside
$G_3$); $\Theta(\rho)$ is a smooth non-negative function on the
half-axis $0\leqslant\rho<+\infty$  that equals 1 as
$0\leqslant\rho\leqslant\delta$ and vanishes as
$\rho\geqslant2\delta$ ($\delta$ being a fixed small positive
number). Thus $\chi_{3, \varepsilon}$ is defined on the whole
waveguide $G(\varepsilon)$ as well as the function $\chi_{3,\,
\varepsilon}v_3$ in (\ref{asympt of wave function}).

Being substituted to (\ref{2D problem}), the function $\chi_{3, \,
\varepsilon}v_3$ gives a discrepancy in the right-hand side of the
Helmholtz equation; the discrepancy is supported near the second
narrow (to the right of it). We compensate the principal part of the
discrepancy by means of the second kind limit problem in the domain
$\Omega_2$. Namely, the discrepancy is rewritten into coordinates
$(\xi_2,\eta_2)$ in $\Omega_2$ and is taken as a right-hand side for
the Laplace equation. The solution $w_2$ of the corresponding
problem (\ref{limit problems 2}) has to be rewritten into
coordinates $(x_2,y_2)$ and multiplied by a cut-off function. As a
result, there arises  the term
$\Theta(r_2)w_2(\varepsilon^{-1}x_2,\varepsilon^{-1}y_2;\varepsilon)$
in (\ref{asympt of wave function}).

Now we substitute the sum of two obtained terms into (\ref{2D
problem}). The principal part of the corresponding discrepancy is
supported in $G_2$ near the second narrow. We compensate it by
solving the problem (\ref{limit problems 1}) in $G_2$ and obtain the
term $\chi_{2, \, \varepsilon}(x,y)v_2(x,y;\varepsilon)$ with
$$
\chi_{2, \, \varepsilon}(x,y)=
\left(1-\Theta(\varepsilon^{-1}r_1)-\Theta(\varepsilon^{-1}r_2)\right)\mathbf{1}_{G_2}(x,y).
$$
Then in a similar way there arise
$$\Theta(r_1)w_1(\varepsilon^{-1}x_1,\varepsilon^{-1}y_1;\varepsilon)
\,\,{\rm and}\,\, \chi_{1,
\,\varepsilon_1}(x,y)v_1(x,y;\varepsilon).$$ At the last step, we
find the function $v_1$ that satisfies both the limit problem
(\ref{limit problems 1}) in $G_1$ and the radiation condition
$$
v_1(x,y;\varepsilon)\sim
s_{12}(\varepsilon)\alpha(\varepsilon)e^{i\nu_1x}\Psi_1(y)+
s_{12}(\varepsilon)\beta(\varepsilon)e^{-i\nu_1x}\Psi_1(y)
$$
as $x\rightarrow-\infty$. The coefficients $\alpha$, $\beta$ and the
entries $s_{11}$, $s_{12}$ of the scattering matrix turn out to be
uniquely determined by a relation between $\alpha$ and $\beta$ that
assures compensation of the principal part of the discrepancy
arising in the problem in $G_1$, and by requirements
$$
s_{12}(\varepsilon)\alpha(\varepsilon)=1,\quad
s_{12}(\varepsilon)\beta(\varepsilon)=s_{11}(\varepsilon).
$$
The remainder $R(x,y;\varepsilon)$ is small in comparison with the
principal part of (\ref{asympt of wave function}) as
$\varepsilon\rightarrow0$.

We specify (\ref{asympt of wave function}) provided $k^2$ varies in
an interval containing a unique simple eigenvalue $k^2_0$ of the
problem (\ref{limit problems 1}) in $G_2$.

\begin{enumerate}
    \item Introduce a special solution $\textbf{v}_3$ of the problem (\ref{limit problems
    1}) in $G_3$ satisfying
    $$
    \textbf{v}_3(x,y)\sim (r_2^{-\pi/\omega}+ar_2^{\pi/\omega})\Phi(\varphi_2)\qquad
    \hbox{as}\;r_2\rightarrow 0
    $$
    (here and below, $(r_j,\varphi_j)$ are polar coordinates with center at
    $O_j$, $j=1,2$; $\Phi(\varphi)=\cos(\pi\varphi/\omega)$), and
    $$
    \textbf{v}_3(x,y)\sim A e^{i\nu_1x}\Psi_1(y)
    \qquad\hbox{as}\;x\rightarrow+\infty.
    $$
    These conditions define $\textbf{v}_3$ uniquely. The constants
    $a$, $A$ (depending on $k$ and on the geometry of $G_3$) have to be
    calculated. We have
    $$
    v_3(x,y;\varepsilon)=\frac{s_{12}(\varepsilon)}{A}\textbf{v}_3(x,y).
    $$
    \item Consider a solution $w_r$ of the homogeneous problem (\ref{limit problems
    2}) satisfying
    \begin{equation}\label{wr}
w_r(\xi,\eta)= \left\{%
\begin{array}{ll}
    (\rho^{\pi/\omega}+\alpha\rho^{-\pi/\omega})\Phi(\varphi)+O(\rho^{-3\pi/\omega}), & \hbox{as $\rho\rightarrow\infty$, $\xi>0$;} \\
    \beta\rho^{-\pi/\omega}\Phi(\pi-\varphi)+O(\rho^{-3\pi/\omega}), & \hbox{as $\rho\rightarrow\infty$, $\xi<0$.} \\
\end{array}%
\right.
\end{equation}
    The constants $\alpha$, $\beta$ (depending on $\Omega$) have to
    be calculated. One can prove that $\beta\neq 0$ (cf.
    \cite{BNPS}, proof of Proposition 3.4). We put
    \begin{eqnarray*}
      \textbf{w}^-(\xi,\eta)&=&\frac{1}{\beta}\left(w_r(\xi,\eta)
      -\zeta_r(\xi,\eta)(\rho^{\pi/\omega}+\alpha\rho^{-\pi/\omega})\Phi(\varphi)
      -\zeta_l(\xi,\eta)\beta\rho^{-\pi/\omega}\Phi(\pi-\varphi)\right),
    \end{eqnarray*}
    where $\zeta_r$ is a cut-off function equal to $1-\Theta$ as
    $\xi>0$ and 0 as $\xi<0$,
    $\zeta_l(\xi,\eta)=\zeta_r(-\xi,\eta)$. We also put
    $$
    \textbf{w}^+(\xi,\eta)=\beta\textbf{w}^-(-\xi,\eta)-\alpha\textbf{w}^-(\xi,\eta).
    $$
    Then
    $$
    w_2(\xi_2,\eta_2;\varepsilon)=\frac{s_{12}(\varepsilon)}{A}
    \left(\varepsilon^{-\pi/\omega}\textbf{w}^-(\xi_2,\eta_2)+
    a\varepsilon^{\pi/\omega}\textbf{w}^+(\xi_2,\eta_2)\right).
    $$
    \item Remind that $k^2_0$ is a simple eigenvalue. Let $v_0$ be an eigenfunction
    corresponding to $k^2_0$ and normalized by
    $\int_{G_2}|v_0|^2dx\,dy=1$. We have
    \begin{equation}\label{v0}
v_0(x,y)\sim \left\{%
\begin{array}{ll}
    b_1r_1^{\pi/\omega}\Phi(\varphi_1), & \hbox{as $r_1\rightarrow 0$;} \\
    b_2r_2^{\pi/\omega}\Phi(\pi-\varphi_2), & \hbox{as $r_2\rightarrow 0$.} \\
\end{array}%
\right.
\end{equation}
    In what follows we assume that $b_1 \neq 0$; such an assumption is fulfilled, for example, if
    $k^2_0$ is the first eigenvalue of problem (\ref{limit problems 1}).
    Since $G_2$ is invariant with respect
    to the transformation $(x,y)\mapsto(d-x,y)$, while
    $d=|x^0_1-x^0_2|$ is the distance between $O_1$ and $O_2$, one
    can prove that $q:=b_2/b_1=\pm1$. Introduce special solutions
    $\textbf{v}_{21}$, $\textbf{v}_{22}$ of the problem (\ref{limit problems 1}) in $G_2$ satisfying
    \begin{eqnarray*}
      \textbf{v}_{21}(x,y)&\sim& \left\{%
\begin{array}{ll}
    \left((k^2-k_0^2)r_1^{-\pi/\omega}+c_1 r_1^{\pi/\omega}\right)\Phi(\varphi_1), & \hbox{as $r_1\rightarrow 0$;} \\
    c_2r_2^{\pi/\omega}\Phi(\pi-\varphi_2), & \hbox{as $r_2\rightarrow 0$,} \\
\end{array}%
\right. \\
      \textbf{v}_{22}(x,y)&\sim& \left\{%
\begin{array}{ll}
    \left(b_2r_1^{-\pi/\omega}+d_1 r_1^{\pi/\omega}\right)\Phi(\varphi_1), & \hbox{as $r_1\rightarrow 0$;} \\
    \left(-b_1r_2^{-\pi/\omega}+d_2 r_2^{\pi/\omega}\right)\Phi(\pi-\varphi_2), & \hbox{as $r_2\rightarrow 0$.} \\
\end{array}%
\right.
    \end{eqnarray*}
    The coefficients $c_1$, $c_2$, $d_1$, $d_2$ depend on $k$ and on
    $G_2$. One can prove that $c_j(k_0)=b_1b_j$. Then
    $$
    v_2(x,y;\varepsilon)=\frac{s_{12}(\varepsilon)}{b_1c_2}
    \bigl((b_1\gamma(\varepsilon)-d_2\delta(\varepsilon))\textbf{v}_{21}(x,y)+c_2\delta(\varepsilon)\textbf{v}_{22}(x,y)\bigr),
    $$
    where
    \begin{equation}\label{gammadelta}
    \gamma(\varepsilon)=\frac{1}{A\beta}\left(\varepsilon^{-2\pi/\omega}-a\alpha\right),\qquad
    \delta(\varepsilon)=-\frac{1}{A\beta}\left(\alpha+a(\beta^2-\alpha^2)\varepsilon^{2\pi/\omega}\right).
\end{equation}
     \item We have
     $$w_1(\xi_1,\eta_1;\varepsilon)=\dfrac{1}{|A|^2}\left((\overline{A}s_{11}(\varepsilon)+A)\varepsilon^{-\pi/\omega}\textbf{w}^-(\xi_1,\eta_1)
    +(a\overline{A}s_{11}(\varepsilon)+
    \overline{a}A)\varepsilon^{\pi/\omega}\textbf{w}^+(\xi_1,\eta_1)\right),$$
     where $s_{11}$ is defined by (\ref{s11}) below.
    \item Introduce a special
     solution $\textbf{v}_1$ of the problem (\ref{limit problems
    1}) in $G_1$ by
    $\textbf{v}_1(x_1,y_1)=\textbf{v}_3(d-x_1,y_1)$. Then
    $$
    v_1(x,y;\varepsilon)=\frac{s_{11}(\varepsilon)}{A}\textbf{v}_1(x,y)+\frac{1}{\overline{A}}\overline{\textbf{v}}_1(x,y),
    $$
    where
    \begin{eqnarray}\nonumber
      s_{11}(\varepsilon) &=&(2ib_1c_2)^{-1}\bigl((k^2-k^2_0)b_1|\gamma(\varepsilon)|^2-
      ((k^2-k^2_0)d_2-b_2c_2)\overline{\gamma(\varepsilon)}\delta(\varepsilon)\\
      &&\label{s11}+
      b_1c_1\gamma(\varepsilon)\overline{\delta(\varepsilon)}-
      (c_1d_2-c_2d_1)|\delta(\varepsilon)|^2\bigr)s_{12}(\varepsilon),
      \\\nonumber
      s_{12}(\varepsilon) &=& 2ib_1c_2\bigl(-(k^2-k^2_0)b_1\gamma(\varepsilon)^2+
      ((k^2-k^2_0)d_2-b_1c_1-b_2c_2)\gamma(\varepsilon)\delta(\varepsilon)\\\label{s12}
      &&+
      (c_1d_2-c_2d_1)\delta(\varepsilon)^2\bigr)^{-1},
    \end{eqnarray}
 while   $\gamma(\varepsilon)$, $\delta(\varepsilon)$ are defined by
    (\ref{gammadelta}). One can verify that $|s_{11}|^2+|s_{12}|^2=1$ (cf. \cite{BNPS}).
\end{enumerate}
Analysis of (\ref{s12}) shows that $T=T(k,\varepsilon)=|s_{12}|^2$
has a sharp peak at $k=k_{res}$,
\begin{equation}\label{kres}
    k^2_{res}=k^2_0-2\alpha
    b_1^2\varepsilon^{2\pi/\omega}+O(\varepsilon^{2\pi/\omega+\tau}),
\end{equation}
where $\tau=\min\{\pi/\omega,2-\sigma\}$, $\sigma$ being a small
positive number. Suppose that $k$ varies in a small neighborhood $I$
of $k_{res}$, $I=\{k:|k-k_{res}|\leqslant c
\varepsilon^{2\pi/\omega+p}\}$, $p>0$. Then (\ref{s12}) takes the
form
$$
s_{12}(k,\varepsilon)=\frac{q(A(k_0)/|A(k_0)|)^2}{1-iP\left(\dfrac{k^2-k^2_{res}}{\varepsilon^{4\pi/\omega}}\right)}(1+O(\varepsilon^p)),
$$
where $P=(2b_1^2\beta^2|A(k_0)|^2)^{-1}$. Hence,
\begin{equation}\label{T}
T(k,\varepsilon)=\frac{1}{1+P^2\left(\dfrac{k^2-k^2_{res}}{\varepsilon^{4\pi/\omega}}\right)^2}(1+O(\varepsilon^p)).
\end{equation}
The width of the peak at its half-height (the so-called a resonator
quality factor) is
\begin{equation}\label{quality factor}
\Upsilon(\varepsilon)=\frac{2}{P}\varepsilon^{4\pi/\omega}.
\end{equation}

\section{Problems and methods for numerical analysis}

The principal parts of asymptotic formulas (\ref{kres}) --
(\ref{quality factor}) for the main characteristics of resonant
tunneling contain the constants $b_1$, $|A|$, $\alpha$, $\beta$. To
find the constants we have to solve numerically several boundary
value problems.  In this section, we state the problems and describe
a way to solve them.  We also outline a method for computing the
waveguide scattering matrix $S$.

To find $b_1$, we solve the spectral problem (\ref{limit problems
1}) in $G_2$ by FEM as usual. Let $V_0$ be an eigenfunction
corresponding to $k_0^2$ and normalized by $\int_{G_2}|V_0(x,
y)|^2\,dxdy=1$ . Then $b_1$ in (\ref{v0}) can be defined by
$$
b_1=\varepsilon^{-\pi/\omega}\frac{V_0(\varepsilon,0)}{\Phi(0)}=
\sqrt{\pi}\varepsilon^{-\pi/\omega}V_0(\varepsilon,0).
$$

Let us calculate $|A|$. In order to avoid dealing with
$\textbf{v}_1$, which increases at $O_1$, we introduce
$\textbf{v}=(\textbf{v}_1-\overline{\textbf{v}}_1)/A$,
\begin{equation}\label{v}
    \textbf{v}(x_1,y_1)=\left\{%
\begin{array}{ll}
    \textbf{a}r_1^{\pi/\omega}\Phi(\varphi_1)
    &\hbox{as}\;r_2\rightarrow 0; \\
    \left(e^{-i\nu_1x_1}+\dfrac{\overline{A}}{A}e^{i\nu_1x_1}\right)\Psi_1(y_1)+O(e^{-\delta|x_1|})
    &\hbox{as}\;x_1\rightarrow-\infty, \\
\end{array}%
\right.
\end{equation}
where $\textbf{a}=2i\hbox{Im}\,a/A$. According to Lemma 4.1 in
\cite{BNPS}, $\hbox{Im}\,a=|A|^2$, so $\textbf{a}=2i\overline{A}$.
Thus, it suffices to calculate $\textbf{a}$. Denote the truncated
domain $G_1\cap\{(x_1,y_1):x_1>-R\}$ by $G_1^R$ and the artificial
part of the boundary $\partial G_1^R\cap\{(x_1,y_1):x_1=-R\}$ by
$\Gamma^R$. Let $V$ be a solution of the problem
\begin{equation}\label{problem3}
    \begin{array}{rll}
\Delta V(x_1,y_1)+k^2V(x_1,y_1)&=0,&(x_1,y_1)\in G_1^R;\\
V(x_1,y_1)&=0,&(x_1,y_1)\in\partial G_1^R\backslash\Gamma^R;\\
\partial_n V(x_1,y_1)+i\nu_1 V
(x_1,y_1) &= 2i\nu_1 e^{i\nu_1R}\Psi_1(y_1), &(x_1,y_1)\in\Gamma^R.
\end{array}
\end{equation}
We find $V$ with FEM and put
$$
\textbf{a}=\sqrt{\pi}\varepsilon^{-\pi/\omega}V(-\varepsilon,0).
$$

Pass to description of a boundary value problem for calculating
$\alpha$, $\beta$ in (\ref{wr}). Denote
$\Omega\cap\{(r,\varphi):r<R\}$ by $\Omega^R$ and
$\partial\Omega\cap\{(r,\varphi):r=R\}$ by $\Gamma^R$. Consider
the problem
\begin{equation}\label{problem2}
    \begin{array}{rll}
\Delta w(\xi,\eta)&=0,&(\xi,\eta)\in\Omega^R;\\
w(\xi,\eta)&=0,&(\xi,\eta)\in\partial \Omega^R\backslash\Gamma^R;\\
\partial_n w(\xi,\eta)+\zeta w(\xi,\eta) &= g(\xi,\eta), &(\xi,\eta)\in\Gamma^R.
\end{array}
\end{equation}
If $w$ is a solution  and $\zeta>0$, then
\begin{equation}\label{est for w}
    \|w;L_2(\Gamma^R)\|\leqslant \zeta^{-1}\|g;L_2(\Gamma^R)\|.
\end{equation}
Indeed, substitute $u=v=w$ to the Green formula
\begin{eqnarray*}
&&(\triangle u,v)_{\Omega^R}=
(\partial_nu,v)_{\partial\Omega^R}-(\nabla u,\nabla
v)_{\Omega^R}\\
&&=(\partial_nu,v)_{\partial\Omega^R\setminus\Gamma^R}+(\partial_nu+\zeta
u,v)_{\Gamma^R}-\zeta(u,v)_{\Gamma^R}-(\nabla u,\nabla
v)_{\Omega^R}
\end{eqnarray*}
and get
$$
0=(g,w)_{\Gamma^R}-\zeta\|w;L_2(\Gamma^R)\|^2-\|\nabla
w;L_2(\Omega^R)\|^2.
$$
From this and the obvious chain of inequalities
$$
\zeta\|w;L_2(\Gamma^R)\|^2\leqslant
\zeta\|w;L_2(\Gamma^R)\|^2+\|\nabla
w;L_2(\Omega^R)\|^2=(g,w)_{\Gamma^R}\leqslant\|w;L_2(\Gamma^R)\|
\,\|g;L_2(\Gamma^R)\|
$$
we obtain (\ref{est for w}). Denote the left part of $\Gamma^R$ by
$\Gamma^R_-$ and the right part of $\Gamma^R$ by $\Gamma^R_+$. Let
$W$ be the solution of (\ref{problem2}) as $\zeta=\pi/\omega R$,
$g|_{\Gamma^R_-}=0$,
$g|_{\Gamma^R_+}=(2\pi/\omega)R^{(\pi/\omega)-1}\Phi(\varphi)$.
Since the asymptotics (\ref{wr}) can be differentiated, $w_r-W$
satisfies (\ref{problem2}) with $g=O(R^{-(3\pi/\omega)-1})$.
According to (\ref{est for w}),
$$
\|w_r-W;L_2(\Gamma^R)\|\leqslant c \frac{\omega
R}{\pi}R^{-(3\pi/\omega)-1}=c'R^{-3\pi/\omega}
$$
as $R\rightarrow+\infty$. We find $W$ with FEM and take
$$
\beta=\frac{W(-R,0)}{\Phi(0)}R^{\pi/\omega}=\sqrt{\pi}W(-R,0)R^{\pi/\omega}.
$$
Obviously,
$\|(w_r-R^{\pi/\omega}\Phi(\varphi))-(W-R^{\pi/\omega}\Phi(\varphi));L_2(\Gamma^R)\|\leqslant
c'R^{-3\pi/\omega}$, therefore we put
$$
\alpha=\frac{W(R,0)-R^{\pi/\omega}\Phi(0)}{\Phi(0)}R^{\pi/\omega}=\sqrt{\pi}W(R,0)R^{\pi/\omega}-R^{2\pi/\omega}.
$$

Finally, we outline the method of calculating the scattering
matrix. Introduce the notation
$$
\begin{array}{rll}
&G(\varepsilon,R)=G(\varepsilon)\cap\{(x,y):-R<x<d+R\},\\
&\Gamma^R_1=\partial G(\varepsilon,R)\cap\{(x,y):x=-R\},\quad
\Gamma^R_2=\partial G(\varepsilon,R)\cap\{(x,y):x=d+R\}
\end{array}
$$
for large  $R$. We search the row $(s_{m1},\ldots,s_{m,2M})$ of
the scattering matrix $s=s(k)$ defined by (\ref{um}),
$m=1,\ldots,M$. As approximation to the row we take the minimizer
of a quadratic functional.  To construct such a functional we
consider the problem
\begin{eqnarray}\label{problem in GR for X}\nonumber
  \Delta\mathcal{X}_m^R+k^2\mathcal{X}_m^R &=& 0\quad \hbox{in }G(\varepsilon,R), \\\nonumber
  \mathcal{X}_m^R &=& 0\quad \hbox{on } \partial
    G(\varepsilon,R)\setminus(\Gamma^R_1\cup\Gamma^R_2),\\\nonumber
  (\partial_n+i\zeta)\mathcal{X}_m^R &=&
  i(-\nu_m+\zeta)e^{-i\nu_mR}\Psi_m(y) + \displaystyle\sum_{j=1}^Ma_j\,i(\nu_j+\zeta)e^{i\nu_jR}\Psi_j(y)\,\hbox{ on
  }\Gamma^R_1,\\\label{problem for X}
  (\partial_n+i\zeta)\mathcal{X}_m^R &=&
  \displaystyle\sum_{j=1}^Ma_{M+j}\,i(\nu_j+\zeta)e^{i\nu_j(d+R)}\Psi_j(y)\,\hbox{ on
  }\Gamma^R_2,
\end{eqnarray}
where $\zeta\in\mathbb{R}\setminus\{0\}$ is an arbitrary fixed
number, and $a_1,\ldots,a_M$ are complex numbers. The solution
$u_m$ to the homogeneous problem (\ref{2D problem}) satisfies the
first two equations (\ref{problem in GR for X}). The asymptotics
(\ref{um}) can be differentiated so $u_m$ satisfies the last two
equations in (\ref{problem for X}) up to  an exponentially small
discrepancy. As approximation for the row
$(s_{m1},\ldots,s_{m,2M})$ we take the minimizer
$a^0(R)=(a^0_1(R),\ldots,a^0_{2M}(R))$ of the functional
\begin{eqnarray}\nonumber
    J_m^R(a_1,\ldots,a_{2M})&=&\|\mathcal{X}_m^R-e^{-i\nu_mR}\Psi_m-\sum\nolimits_{j=1}^Ma_je^{i\nu_jR}\Psi_j;L_2(\Gamma^R_1)\|^2\\\label{functional}
    &+&\|\mathcal{X}_m^R-\sum\nolimits_{j=1}^Ma_{M+j}\,e^{i\nu_j(d+R)}\Psi_j;L_2(\Gamma^R_2)\|^2,
\end{eqnarray}
where $\mathcal{X}_m^R$ is a solution to problem (\ref{problem in GR
for X}). As shown in \cite{PS}, $a^0_j(R, k)\rightarrow s_{mj}(k)$
with exponential rate as $R\rightarrow\infty$ and $j=1,\ldots,2M$.
To find the dependence of $\mathcal{X}_m^R$ on $a_1,\ldots,a_{2M}$,
we consider the problems
\begin{eqnarray}\nonumber
  \Delta v_j^{\pm} +k^2v_j^{\pm} &=& 0\quad \hbox{in }G(\varepsilon,R), \\\nonumber
  v_j^{\pm} &=& 0\quad \hbox{on } \partial
    G(\varepsilon,R)\setminus(\Gamma^R_1\cup\Gamma^R_2),\\\nonumber
  (\partial_n+i\zeta)v_j^{\pm} &=&
  i(\mp\nu_j+\zeta)e^{\mp i\nu_jR}\Psi_j\quad\hbox{on
  }\Gamma^R_1,\\\label{problem in GR for v+- 1...M}
  (\partial_n+i\zeta)v_j^{\pm} &=& 0\quad\hbox{on
  }\Gamma^R_2;\qquad j=1,\ldots,M;
\end{eqnarray}
and
\begin{eqnarray}\nonumber
  \Delta v_j^{\pm} +k^2v_j^{\pm} &=& 0\quad \hbox{in }G(\varepsilon,R), \\\nonumber
  v_j^{\pm} &=& 0\quad \hbox{on } \partial
    G(\varepsilon,R)\setminus(\Gamma^R_1\cup\Gamma^R_2),\\\nonumber
  (\partial_n+i\zeta)v_j^{\pm} &=& 0\quad\hbox{on
  }\Gamma^R_1,\\\label{problem in GR for v+- M+1...2M}
  (\partial_n+i\zeta)v_j^{\pm} &=&
  i(\mp\nu_j+\zeta)e^{\mp i\nu_j(d+R)}\Psi_j\quad\hbox{on
  }\Gamma^R_2;\qquad j=M+1,\ldots,2M.
\end{eqnarray}
Express $\mathcal{X}_m^R$ by means of the solutions
$v_j^{\pm}=v_{j,R}^{\pm}$ to problems (\ref{problem in GR for v+-
1...M})--(\ref{problem in GR for v+- M+1...2M}). We have
$\mathcal{X}_m^R=v_{m,R}^++\sum_ja_jv_{j,R}^-$. Let us introduce
the $2M\times 2M$--matrices with entries
\begin{equation*}
    \begin{split}
    &\mathcal{E}^{R}_{mj}=\left((v_m^--e^{i\nu_mR}\Psi_m),(v_j^--e^{i\nu_jR}\Psi_j)\right)_{\Gamma^R_1}+\left(v_m^-,v_j^-\right)_{\Gamma^R_2},\\
    &\mathcal{E}^{R}_{m,M+j}=\left((v_m^--e^{i\nu_mR}\Psi_m),v_{M+j}^-\right)_{\Gamma^R_1}+\left(v_m^-,(v_{M+j}^--e^{i\nu_j(d+R)}\Psi_j)\right)_{\Gamma^R_2},\\
    &\mathcal{E}^{R}_{M+m,j}=\left(v_{M+m}^-,(v_j^--e^{i\nu_jR}\Psi_j)\right)_{\Gamma^R_1}+\left((v_{M+m}^--e^{i\nu_m(d+R)}\Psi_m),v_j^-\right)_{\Gamma^R_2},\\
    &\mathcal{E}^{R}_{M+m,M+j}=\left(v_{M+m}^-,v_{M+j}^-\right)_{\Gamma^R_1}+\left((v_{M+m}^--e^{i\nu_m(d+R)}\Psi_m),(v_{M+j}^--e^{i\nu_j(d+R)}\Psi_j)\right)_{\Gamma^R_2};
\end{split}
\end{equation*}
\begin{equation*}
    \begin{split}
    &\mathcal{F}^{R}_{mj}=\left((v_m^+-e^{-i\nu_mR}\Psi_m),(v_j^--e^{i\nu_jR}\Psi_j)\right)_{\Gamma^R_1}+\left(v_m^+,v_j^-\right)_{\Gamma^R_2},\\
    &\mathcal{F}^{R}_{m,M+j}=\left((v_m^+-e^{-i\nu_mR}\Psi_m),v_{M+j}^-)\right)_{\Gamma^R_1}+\left(v_m^+,(v_{M+j}^--e^{i\nu_j(d+R)}\Psi_j)\right)_{\Gamma^R_2},\\
    &\mathcal{F}^{R}_{M+m,j}=\left(v_{M+m}^+,(v_j^--e^{i\nu_jR}\Psi_j)\right)_{\Gamma^R_1}+\left((v_{M+m}^+-e^{-i\nu_m(d+R)}\Psi_m),v_j^-\right)_{\Gamma^R_2},\\
    &\mathcal{F}^{R}_{M+m,M+j}=\left(v_{M+m}^+,v_{M+j}^-\right)_{\Gamma^R_1}+\left((v_{M+m}^+-e^{-i\nu_m(d+R)}\Psi_m),(v_{M+j}^--e^{i\nu_j(d+R)}\Psi_j)\right)_{\Gamma^R_2},
\end{split}
\end{equation*}
$j,m=1,\ldots,M$. We also put
\begin{equation*}
    \begin{split}
    &\mathcal{G}^{R}_{m}=\left((v_m^+-e^{-i\nu_mR}\Psi_m),(v_m^+-e^{-i\nu_jR}\Psi_j)\right)_{\Gamma^R_1}+\left(v_m^+,v_m^+\right)_{\Gamma^R_2},\\
        &\mathcal{G}^{R}_{M+m}=\left(v_{M+m}^+,v_{M+m}^+\right)_{\Gamma^R_1}+\left((v_{M+m}^+-e^{-i\nu_m(d+R)}\Psi_m),(v_{M+m}^+-e^{-i\nu_j(d+R)}\Psi_j)\right)_{\Gamma^R_2},
\end{split}
\end{equation*}
m=1,\ldots,M.
The functional (\ref{functional}) can be written in the form
$$
J_m^R(a, k)=\langle
a\mathcal{E}^R(k),a\rangle+2\mbox{Re}\,\langle\mathcal{F}^R_m(k),a\rangle+\mathcal{G}^R_m(k),
$$
where $\mathcal{F}^R_m$ is the $m$-th row of the matrix
$\mathcal{F}^R$ and $\langle\cdot,\cdot\rangle$ is the inner
product on $\mathbb{C}^M$. The minimizer $a^0 =a^0(R,k)$ (a row)
satisfies $a^0\mathcal{E}^R+\mathcal{F}^R_m=0$. Recall that we are
searching the $m$-th row of the scattering matrix as
$m=1,\ldots,M$. Along the same arguments one can prove that the
found minimizer for $m=M+1,\ldots,2M$ serves as approximation to
the $m$-th row of the scattering matrix. Therefore, as
approximation $s^R(k)$ for the scattering matrix $s(k)$ we take a
solution to the equation $s^R\mathcal{E}^R+\mathcal{F}^R=0$.

When $M=1$, i.e. $k^2$ is between the first and the second
thresholds, we take $\zeta=-\nu_1$. Then $v^-_1=v_2^-=0$,
$\mathcal{E}^R=(1/\nu_1)\hbox{Id}$, and $s^R=-\nu_1\mathcal{F}^R$.

\section{Comparison of asymptotic and numerical results}

\begin{figure}[!htbp]
    \centering
        \includegraphics[scale=.6]{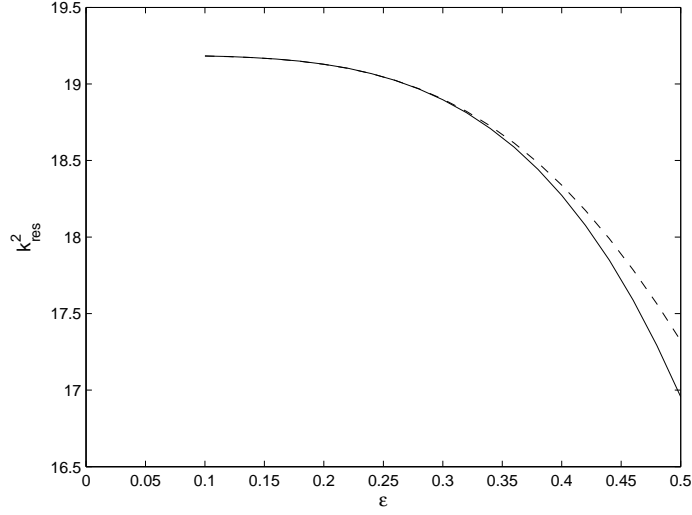}
\caption{Asymptotic description $k^2_{res, a}(\varepsilon)$ (solid
curve) and numerical description $k^2_{res, n}(\varepsilon)$
(dashed curve) for resonant energy $k^2_{res} (\varepsilon)$.}
\label{Fig.k^2}
\end{figure}

Let us compare the asymptotics $k_{res, a}^2 (\varepsilon)$ of
resonant energy $k^2_{res}(\varepsilon)$ and the approximate value
$k^2_{res, n}(\varepsilon)$ obtained by numerical method. Fig. \ref{Fig.k^2}
shows good agreement with the values for $0.1 \leq \varepsilon
\leq 0.5$. We have
$$
|k^2_{res, a}(\varepsilon) - k^2_{res, n}(\varepsilon)|/k^2_{res,
a}(\varepsilon)\leq 10^{-3}
$$
for $0.1\leq \varepsilon \leq 0.3$ and only for $\varepsilon =0.5$
the ratio approaches $2.10^{-2}$. For $\varepsilon < 0.1$ the
numerical method is ill-conditioned.

\begin{figure}[!hp]
    \centering
        \includegraphics[scale=.7]{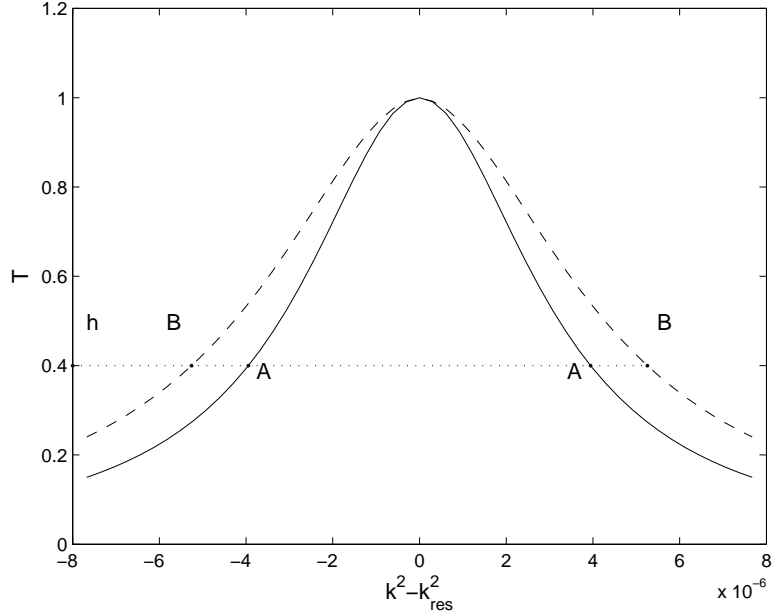}
\caption{Transition coefficient for $\varepsilon =0.2$: asymptotic
description $T_a (k^2 - k^2_{res, a})$ (solid curve) and numerical
description $T_n (k^2 - k^2_{res, n})$ (dashed curve) for
transition coefficient $T (k^2 - k^2_{res})$. The width of
resonant peak at height $h$: asymptotic $\Delta_a (h,
\varepsilon)= AA$;  numerical $\Delta_n(h, \varepsilon)= BB$.}
\label{Fig.peaks}
\end{figure}

\begin{figure}[!htbp]
    \centering
        \includegraphics[scale=.6]{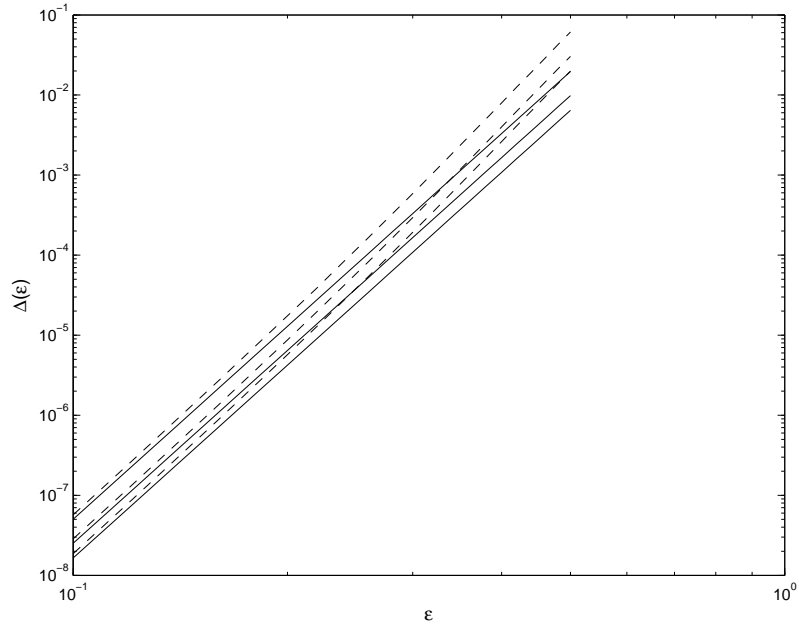}
\caption{The width $\Delta (h, \varepsilon)$ of resonant peak for
various $h$ (dashed line for numerical description, solid line for
asymptotic description): line 1 for $h=0.2$; line 2 for $h=0.5$;
line 3 for $h=0.7$.}
\label{Fig.delta}
\end{figure}

\begin{figure}[!htbp]
    \centering
        \includegraphics[scale=.6]{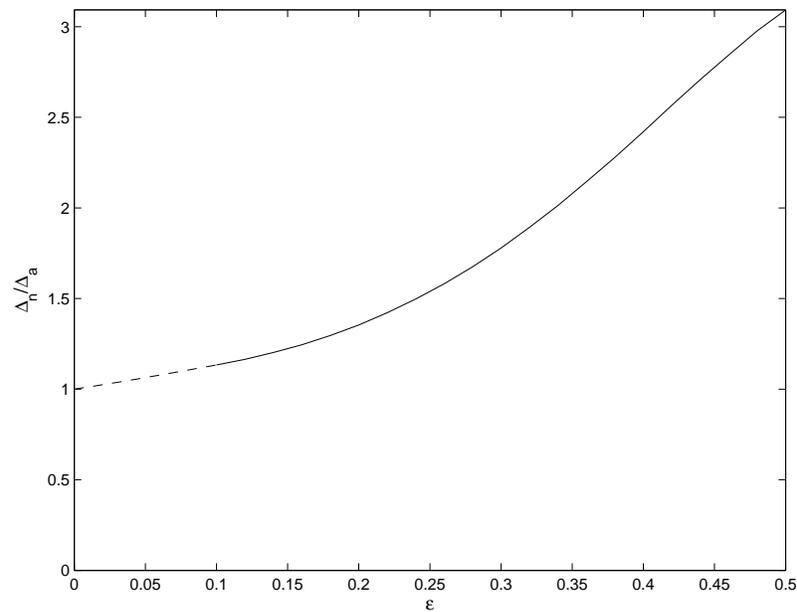}
\caption{Ratio $\Delta_n (h, \varepsilon)/\Delta_a (h,
\varepsilon)$ as function in $\varepsilon$. The ratio is
independent of $h$ within the accuracy of the analysis.}
\label{Fig.dn_da}
\end{figure}

The difference between the asymptotic and numerical values is more
significant for larger $\varepsilon$ because the asymptotics
becomes not reliable. However, as the numerical method shows, for
$\varepsilon \geq 0.5$ the resonant peak turns out to be so wide
that the resonant tunneling phenomenon  dies out by itself.

The forms of "asymptotic"\, and "numerical"\, resonant peaks are
almost the same (see Fig. \ref{Fig.peaks}). The difference between the peaks is
quantitatively depicted in Fig. \ref{Fig.delta}. Moreover, it turns out that
the ratio of the width $\Delta_n (h, \varepsilon)$ of numerical
peak at height $h$ to $\Delta_a (h, \varepsilon)$ of asymptotic
peak is independent of $h$. The ratio as function in $\varepsilon$
is displayed in Fig. \ref{Fig.dn_da}.

Note that for $\varepsilon =0.1$, i.e., at the left end of the
band where the numerical and asymptotic results can be compared,
the disparity of the results is more significant for the width of
resonant peak than that for the resonant energy.


\end{document}